# Microstrip Patch Antenna Design at 10 GHz for X Band Applications

Mehmet Karahan ✉
*Electrical and Electronics Engineering, TOBB University of Economics and Technology, Turkey*

Mertcan Inal
*Electrical and Electronics Engineering, TOBB University of Economics and Technology, Turkey*

Alperen Dilmen
*Electrical and Electronics Engineering, TOBB University of Economics and Technology, Turkey*

Furkan Lacinkaya
*Electrical and Electronics Engineering, TOBB University of Economics and Technology, Turkey*

Ahmet Nuri Akay
*Materials Science and Nanotechnology Engineering, TOBB University of Economics and Technology, Turkey*

Cosku Kasnakoglu
*Electrical and Electronics Engineering, TOBB University of Economics and Technology, Turkey*

**Abstract:**

Microstrip patch antennas are used in satellite imaging systems, wireless communication equipment, military radios, GPS (Global Positioning System) and GSM (Global System for Mobile Communications) applications. Its advantages are its small size and light weight, thin structure, low power consumption, use in dual frequency applications, and patching in various geometric shapes. Developing technology has facilitated and accelerated the production of microstrip antennas. In this study, microstrip antenna design operating at 10 GHz frequency for X band applications has been made. X band is used for air traffic control, weather traffic control, vessel traffic control, defense tracking and vehicle speed detection, terrestrial communications and networking, space communications and amateur radio. HFSS program was used in antenna design. AWR program was used to find transmission line parameters. In addition, MATLAB program was used to calculate some parameters. First of all, information is given about the working principle of the antenna, the selected dielectric layer and the working frequency. Schematic drawings of the designed antenna were made from above and from the side. S11 reflection coefficient magnitude graphs are drawn below and above the operating frequency. The radiation pattern is drawn for the E-plane and H-plane at the operating frequency. 3-D (dimensional) plot of antenna gain at operating frequency is drawn. The simulations performed have shown that the designed antenna works successfully.

# Introduction

Microstrip patch antennas are low profile antennas. They are used in low profile applications at frequencies above 100 MHz (Singh et al., 2011). A metal patch mounted at a ground level with a dielectric material in between creates a microstrip (Deepa et al., 2022). The patch on the upper surface is made of conductive materials such as copper or gold (Bisht et al., 2014). The geometric shape of the conductor to be used may vary according to the design features. Square, rectangle, ellipse, ring etc. can be used in shapes (Shome et al., 2019). Microstrip antennas could be used in different applications such as aircrafts, spacecrafts, satellites, missiles, mobile radios, and wireless communications (Mishra et al., 2016). Microstrip patch antennas can also be used in unmanned aerial vehicles due to its miniaturized dimensions (Karahan et al., 2021). Moreover, microstrip patch antennas are used in video rate imaging systems (Dhillon et al., 2017; Karahan et al., 2022). Also, microstrip patch antennas can be used with a low noise amplifier. In this way, the effect of noise in satellite systems and UAVs is minimized (Kouhalvandi et al., 2017; Karahan et al., 2021).

Rana et al. designed a microstrip patch antenna operating at 2.4 GHz for use in wireless communication (Rana et al., 2022). They used CST studio suite program for the design. They achieved low returns and high gain in their designs. Colaco and Lohani (2020) developed a microstrip patch antenna operating at 26 GHz. Since they need high data and high bandwidth, they have developed a design using 5G millimeter wave bands. They used FEKO software for simulation and analysis. Kiani et al. (2021) designed a microstrip patch antenna using reconfigurable graphene material. They stressed that changes in the chemical potential of graphene are directly caused by changes in Fermi energy. They used graphene material to adjust the polarization of the microstrip patch antenna. They carried out their simulations in the frequency range of 0.65 to 0.7 THz. Hocini et al. (2019) designed a terahertz microstrip patch antenna based on photonic crystals. They obtained best antenna characteristics 0.65 THz. Their proposed antenna had high radiation efficiency as 90.84%. Wang et al. (2023) proposed a frequency reconfigurable antenna design. They designed a frequency reconfigurable microstrip patch antenna using graphane film. They used the graphane film in direct contact with the Si/SiO2 substrate. Their test results showed that center frequency of their antenna was 29.6 GHz when the bias voltage was 0 V and 40 GHz when the bias voltage was 9 V. Ghimire et al. (2023) designed a slot-line-based microstrip patch antenna. They fabricated and analyzed this antenna inside a far-field anechoic chamber. The measured results showed that antenna had a frequency range of 8.5 to 11 GHz. Benlakehal et al. (2023) developed a graphene based microstrip patch antenna in 0.636 THz band. They developed a 1x2 microstrip patch antenna array based on photonic crystals. They obtained a high gain as 11.53 dB.

In this research, a microstrip patch antenna design operating at 10 GHz frequency was carried out. The microstrip patch antenna design was developed for X band applications. A design with high gain, small size, thin structure, light weight and low power consumption has been achieved. A computer program called HFSS was used for antenna design. AWR program was used to obtain transmission line parameters. The MATLAB program was used to make the necessary mathematical calculations. At certain intervals of the antenna's operating frequency, s11 reflection coefficient magnitude graphs were drawn. Antenna's radiation pattern is drawn for the E-plane and H-plane at the operating 10 GHz frequency. 3-D plot of antenna gain at operating 10 GHz frequency is shown. The obtained simulation results proved that the designed microstrip patch antenna works successfully.

## General Specifications of the Microstrip Patch Antenna

Microstrip is a type of electrical transmission line that can be manufactured using printed circuit board technology and is used to transmit microwave-frequency signals. It consists of a conductive strip separated from the ground plane using a dielectric layer known as the substrate (Ihamji et al., 2019). Many microwave circuit elements such as antennas, synchronizers, filters and power dividers can be made using microstrips. Microstrips are cheaper, lighter and more compact than standard waveguides (Varshney et al., 2020). Microstrip patch antenna basically consists of a metal ground layer at the bottom, a substrate consisting of dielectric material in between, and a metal patch at the top that provides radiation (Mishra et al., 2022). A typical microstrip patch antenna is shown in Figure 1.

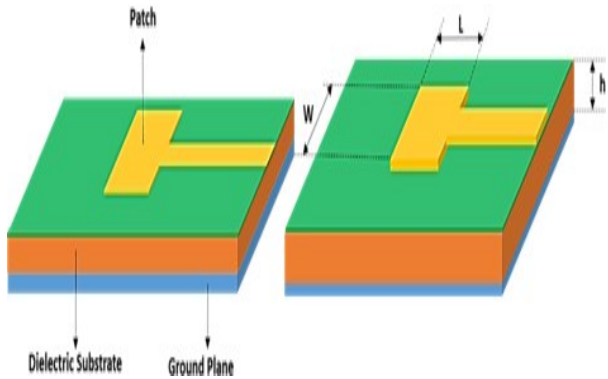

Figure 1. A Typical Microstrip Patch Antenna

The dielectric ground with the patches is not magnetic. The small dielectric constant of the dielectric ground causes the fringe areas to increase, which affects the radiation (Lee et al., 2019). In general, when designing the antenna, it is preferred that the dielectric constant is between 2.2 and 12 (Hashim et al., 2022). The length L, width W and thickness H are effective in characterizing this type of antenna.

## Excitation of the Microstrip Patch Antenna

Transmission line feeding, coaxial cable feeding or inset (embedded) feeding can be used for patch excitation (Ashraf et al., 2023; Kashyap et al., 2022). In this research, inset feeding was used due to space constraints. In this method, $Z_{in}(R)$ is pulled to the desired location by starting from the input impedance ($Z_{in}(0)$) when there is no inset (Yan et al., 2021). Figure 2 shows the schematic representation of inset feeding. Its formula is given in equation 1.

$$Z_{in}(R) = \cos^2(\pi R/L) Z_{in}(0) \qquad (1)$$

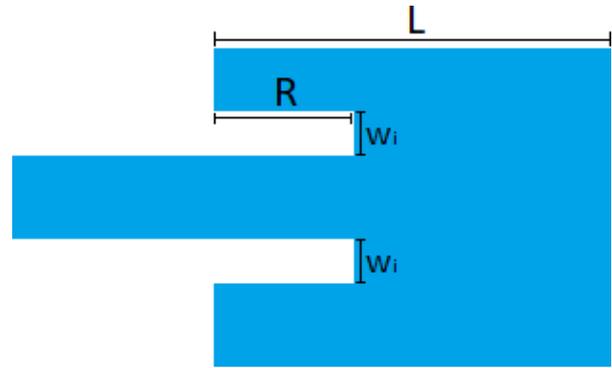

Figure 2. Schematic Representation of Inset Feeding

## Excitation of the Microstrip Patch Antenna

Excitation of the conductive patch, on the other hand, causes an electromagnetic wave movement from the edges of the patch to the ground. Waves reflected from the ground propagate into space. The areas formed on the edges of the conductive patch are called fringing areas and this phenomenon is called fringing effect (Al Ahmad et al.; 2021). Figure 3 gives the schematic representation of fringing areas. The radiation of the antenna occurs as a result of this event. Waves perpendicular to the patch dampen each other and do not radiate, waves fringing from the corners make the radiation.

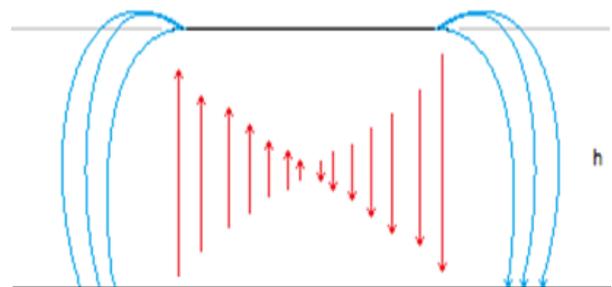

Figure 3. Schematic Representation of Fringing Areas

## Design of the Microstrip Patch Antenna

In this section, the operating frequency of the antenna, the selected dielectric layer, and the design procedures are explained. A side and top schematic view of the designed antenna is given.

## Frequency and Dielectric Layer

It was stated in the design specifications that the communication system uses certain frequencies in the 10-12 GHz range, so 10 GHz within this range was chosen as the center frequency. In the dielectric layer, RO4003 material was chosen because of its high frequency performance, low loss and widespread use in microstrip antenna designs (Khan et al., 2012). The dielectric constant of this material is 3.4 and its tangent loss is 0.002 (Vishnu Chandar et al., 2021).

## Design Procedures

In this research, HFFS program was used for simulation and modeling purposes. AWR program is used to organize some graphs and find transmission line parameters. In addition, MATLAB program was used for some calculations.

First of all, the dimensions of the antenna were determined. The operating frequency of the antenna is determined by L (length). The center frequency is calculated approximately as in equation 2, where c is the speed of light.

$$f_c \approx c/2L\sqrt{\varepsilon_r} = 1/2L\sqrt{\mu_0 \varepsilon_0 \varepsilon_r} \qquad (2)$$

Equation 3 is obtained by subtracting L from equation 2.

$$L \approx c/2f_c\sqrt{\varepsilon_r} \qquad (3)$$

When the equations were solved using the MATLAB program, L = 7.96 was found. Another parameter of the antenna, w (patch width), is determined by the following formula:

$$w = (c/2f_c)/(\sqrt{2}/(\varepsilon_r +1)) \qquad (4)$$

When the equations were solved with the MATLAB program, w = 9.94 mm. Equation 5 was used to calculate the h (height). When this equation was solved with MATLAB, it was found that h = 0.96.

$$(0.0606 \lambda)/(\sqrt{\varepsilon_r}) \qquad (5)$$

As $\varepsilon r$ (dielectric constant) decreases, the effective length of the antenna also changes due to the increase in fringing areas. There may be deviations in $f_c$ (center frequency) due to these changes. Therefore, the antenna effective length ($L_{eff}$), normalized extension in length ($\Delta L$) and effective dielectric constant ($\varepsilon r_{eff}$) are additionally calculated below:

$$\varepsilon r_{eff} = (\varepsilon r +1)/2 + (\varepsilon r -1)/2[1 + 1.2h/w]^{-1/2} \qquad (6)$$

$$\Delta L = 0.142h[(\varepsilon_{reff} + 0.3)(w/h + 0.264)]/[(\varepsilon_{reff} - 0.258)(w/h + 0.8)] \qquad (7)$$

$$L_{eff} = L + \Delta L \qquad (8)$$

When the above equations are solved with MATLAB program, $\varepsilon r_{eff}$ = 3.1417 , $\Delta L$ = 0.4513 $mm$ , $L_{eff}$ = 8.8582 $mm$ results are found.

In inset (embedded) feeding, the following equation was obtained by using the equation (1) and starting from the input impedance ($Z_{in}(0)$)= 204.75$\Omega$.

$$R = \cos^{-1}\left(\frac{Z_{in}(R)}{Z_{in}(0)}\right)\frac{L}{\pi} \qquad (9)$$

Then, using the MATLAB program, R= 2.6689 mm was calculated. The width value at the embedded feed was calculated as w = 0.3313 mm. For the design of the microstrip line, parameters such as line length, line width, line height were found by using the Microstrip section of the AWR program. These parameters are shown in Figure 4.

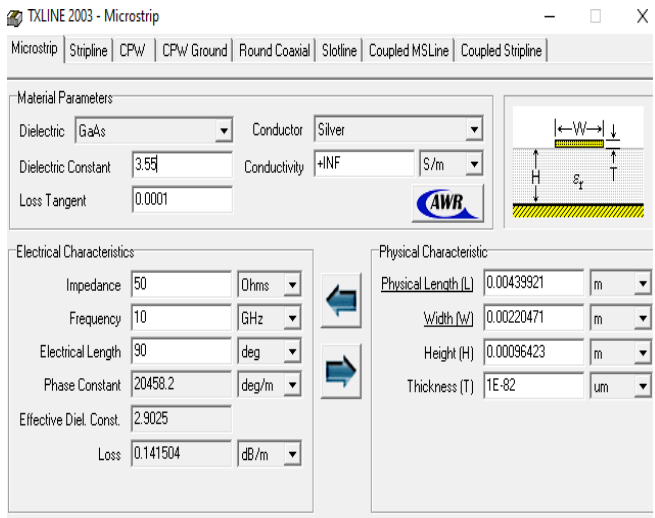

**Figure 4. Calculation of Microstrip Line Parameters in AWR**

The parameters of the designed antenna are given in Table 1.

**Table 1. Design Parameters of the 10 GHz X Band Microstrip Patch Antenna**

| Parameter | Value |
|---|---|
| Working frequency | 10 GHz |
| Wavelength | 3 cm |
| Antenna width | 1.6 cm |
| Antenna length | 1.6 cm |
| Antenna effective length | 0.88582 cm |
| Antenna height | 0.0964 cm |
| Patch length | 0.7554 cm |
| Patch width | 0.9938 cm |
| Patch height | neglected |
| Dielectric base material | RO4003 |
| Dielectric base dielectric constant | 3.4 |
| Effective dielectric constant | 3.1417 |
| Dielectric base tangent loss | 0.002 |
| Dielectric base height | 0.0964 cm |
| Dielectric base length | 1.6 cm |
| Dielectric base width | 1.6 cm |
| Band width | 0.2832 GHz |
| Gain | 6.8 dB |
| Impedance | 64 Ω |
| Transmission line length | 4.399 cm |
| Transmission line width | 2.205 cm |
| Transmission line thickness | neglected |
| Transmission line impedance | 50 Ω |
| Embedded feed w | 0.03313 cm |
| Embedded feed R | 0.2669 cm |

## Schematics of the Designed Antenna

The designed antenna is shown schematically in Figure 5, showing the design parameters and dimensions. It is seen that the total volume rule (1.6 cm x 1.6 cm x 1 mm) given in the design specifications is followed here. Of the values calculated in Section 3, all except L remained the same. The reason for the change of L is the change of $L_{eff}$ due to the fringing areas, as emphasized earlier. L was found by modifying it with the HFSS program to provide the center frequency as its graph is given in the next sections. The patch and ground plane parts shown in the figure are taken as PEC (perfect electrical conductor), and the thickness of the conductive surfaces is neglected.

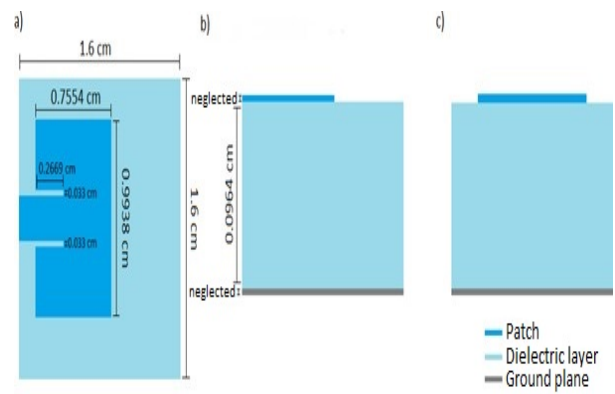

**Figure 5. Schematic Views of the Designed Antenna. a) XY (above) view b) YZ (side) view c) XZ (side) view.**

## Simulations

In this section, s11 reflection coefficient magnitude graphs, antenna input impedance graphs, radiation patterns for E plane and H plane and antenna gain graphs are drawn.

## Simulations of the S11 Reflection Characteristics

S11 is a measure of how much power is reflected back at the antenna port due to mismatch from the transmission line (Iqbal et al., 2021). When $S11$ is plotted between 500 MHz below and above the determined operating frequency of 10 GHz, as seen in Figure 6, the operating frequency of the antenna has changed due to the fringing areas. Fringing areas cause the effective length to change as mentioned before. Therefore, in HFSS, L length was manually changed and an L providing 10 GHz was obtained. Figure 7 describes the s11 reflection coefficient magnitude graph according to the frequency in the range 500 MHz below and above the operating frequency for the manually found L. After obtaining Figure 7, the bandwidth at -10 dB is found as in equation 10.

$$f_2 - f_1 = (10.1307 - 9.8475)\ GHz = 0.2832\ GHz \quad (10)$$

The bandwidth was calculated as in equation 11. This complies with the requirement of design specifications that the -10dB bandwidth (BW) of the desired antenna should be at least 1.6%.

$$BW\% = [(f_2 - f_1)/10]100 = 2.83\% \quad (11)$$

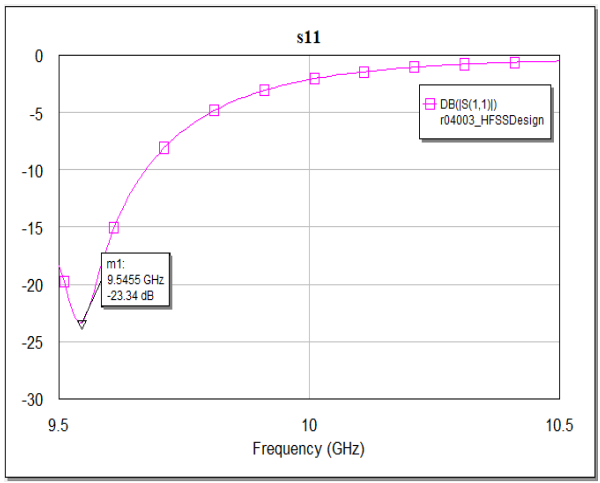

Figure 6. Reflection Coefficient (*S*11) Simulation

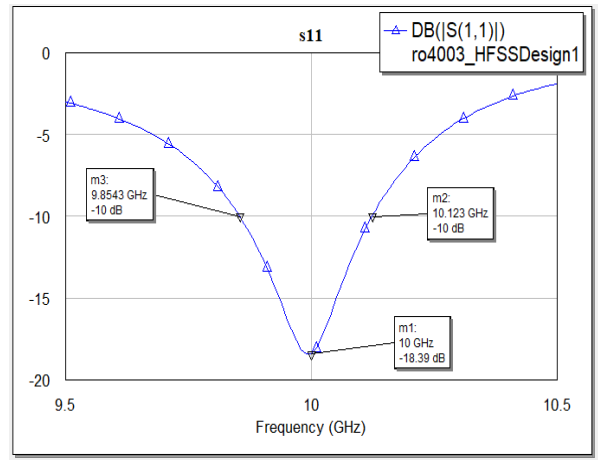

Figure 7. Reflection Coefficient (*S*11) Simulation for the Manually Found L Value

## Antenna input impedance simulations

In Figure 8, the real graph, the imaginal graph and the magnitude graph of the antenna input impedance are plotted between 9.75 GHz and 10.25 GHz. As can be seen, at 10 GHz, the real impedance is 64$\Omega$ and the imaginal impedance is very close to zero. In general, it can be seen that the antenna input impedance is around 50$\Omega$ between 9.75-10.15 GHz.

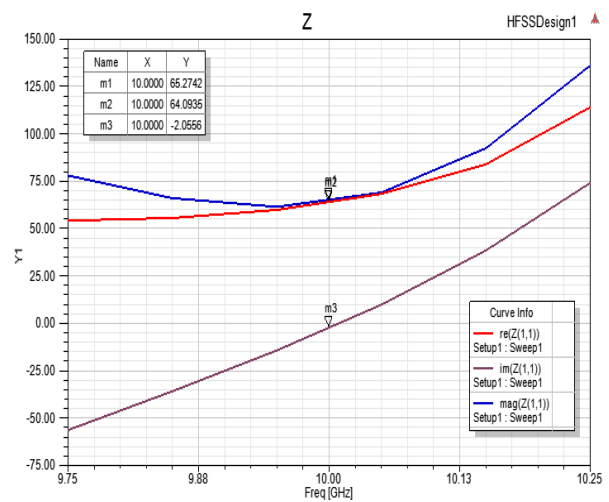

Figure 8. Antenna Input Impedance Graph Between 9.75 GHz and 10.25 GHz Range

Figure 9 shows the impedance graph of the antenna's input port. For 9.75 GHz and 10.25 GHz, the real impedance is 50$\Omega$ and the imaginal impedance is 0 at all frequency values.

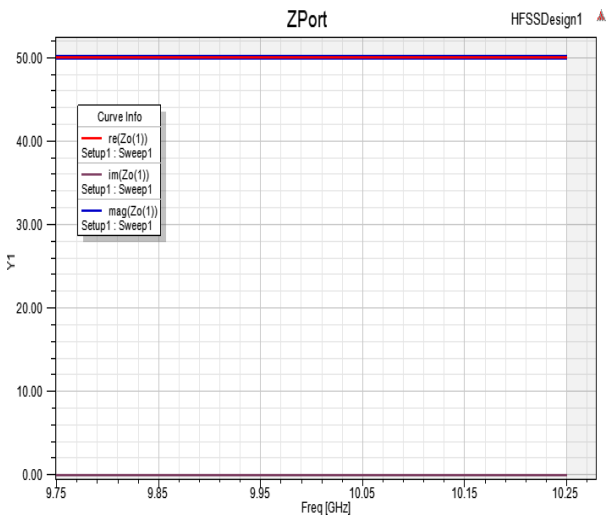

Figure 9. Antenna Input Impedance Graph Between 9.75 GHz and 10.25 GHz Range

## Simulations of the radiation pattern for E-Plane and H-Plane at 10 GHz

Considering the direction of the electric field and the radiation direction, the E plane is the YZ plane, i.e. $\emptyset = \pi/2$ plane, and the H plane is the $\emptyset = 0$ plane. Considering these, radiation patterns at 10 GHz are drawn for E and H planes, respectively, in Figure 10 and Figure 11.

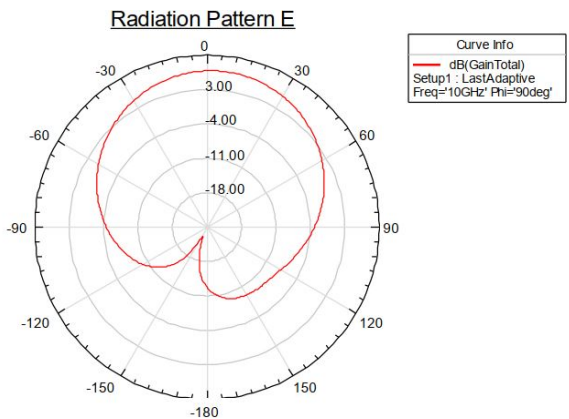

Figure 10. Radiation Pattern for E Plane at Operating Frequency

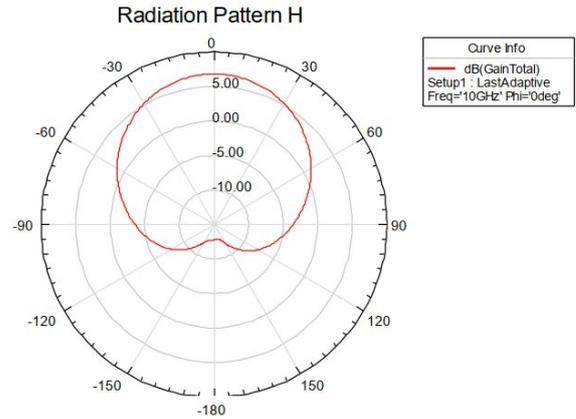

Figure 11. Radiation Pattern for H Plane at Operating Frequency

## Antenna Gain Simulations

Antenna gain at 10 GHz is plotted in 3D in Figure 12. As can be seen from the Figure 12, the antenna gain is higher than 5 dB.

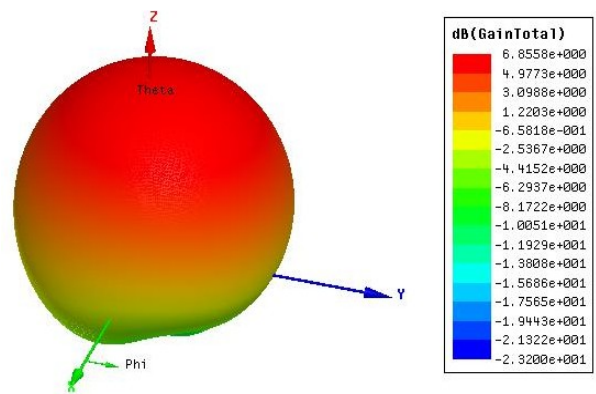

Figure 12. 3D Antenna Gains at Operating Frequency

Antenna gain in the 9.75 and 10.25 GHz range is plotted in Figure 13 depending on $\theta$. As can be seen, the antenna gain is equal to 6.8 dB when $\theta = 0$.

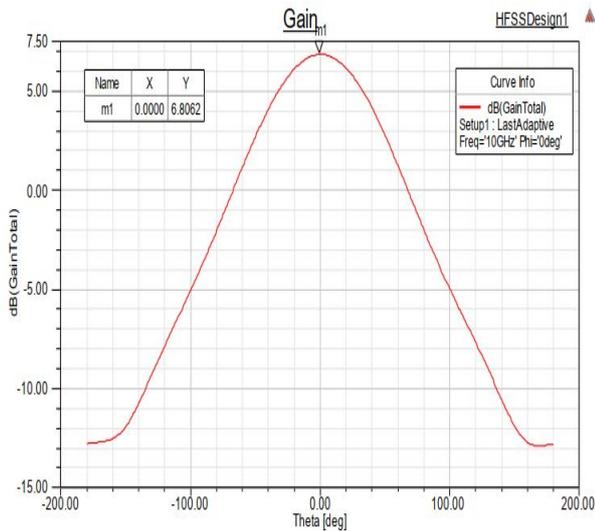

Figure 13. The Graph of Antenna Gain Connected to θ

## Conclusion

In this study, the design of a microstrip patch antenna at 10 GHz frequency for X band applications is explained. First of all, the usage areas, structure and working principles of the microstrip patch antenna are explained. HFSS, AWR and MATLAB programs were used in the antenna design. The equations used in this design are explained one by one. Using MATLAB program, these equations were solved and the values of the parameters were found. The schematic drawings of the antenna are given from the top and from the side. In the simulation section, S11 characteristic graphics, input impedance graphics, E and H plane radiation patterns and antenna gain graphics were drawn. The parameters used in the antenna design are presented in a table. Simulation results show that the antenna works as desired and meets the X Band design criteria.

WBNs and ISM Applications: A Review WBNs and ISM Applications: A Review. *Electronics*, vol. 11(15),pp.2470. https://doi.org/10.3390/electronics11152470

Hocini, A., Temmar, M. N., Khedrouche, D., & Zamani, M. (2019). Novel approach for the design and analysis of a terahertz microstrip patch antenna based on photonic crystals. *Photonics and Nanostructures-Fundamentals and Applications*,36,100723. https://doi.org/10.1016/j.photonics.2019.100723

Ihamji, M., Abdelmounim, E., Bennis, H., & Latrach, M. (2019). Design of miniaturized antenna for RFID applications. In *Emerging Innovations in Microwave and Antenna Engineering*, (pp.325-362). IGI Global. http://dx.doi.org/10.4018/978-1-5225-7539-9.ch010

Iqbal, A., Al-Hasan, M., Mabrouk, I. B., Basir, A., Nedil, M., & Yoo, H. (2021). Biotelemetry and wireless powering of biomedical implants using a rectifier integrated self-diplexing implantable antenna. *IEEE Transactions on Microwave Theory and Techniques*, vol. 69(7), pp. 3438-3451. https://doi.org/10.1109/tmtt.2021.3065560

Karahan, M., Akay, A. N., & Kasnakoglu, C. (2021). Nonlinear modeling and robust control of a quadrotor uav under uncertain parameters and white gaussian noise. In *2021 5th International Symposium on Multidisciplinary Studies and Innovative Technologies (ISMSIT)* (pp. 252-256). IEEE. https://doi.org/10.1109/ISMSIT52890.2021.9604685

Karahan, M., & Kasnakoglu, C. (2021). Modeling a Quadrotor Unmanned Aerial Vehicle and robustness analysis of different controller designs under parameter uncertainty and noise disturbance. *Control Engineering and Applied Informatics*, vol. 23(4), pp. 13-24.

Karahan, M., Lacinkaya, F., Erdonmez, K., Eminagaoglu, E. D., & Kasnakoglu, C. (2022). Face detection and facial feature extraction with machine learning. In *Intelligent and Fuzzy Techniques for Emerging Conditions and Digital Transformation: Proceedings of the INFUS 2021 Conference, held August 24-26, 2021. Volume 2* (pp. 205-213). Springer International Publishing. https://doi.org/10.1007/978-3-030-85577-2_24

Kashyap, N., Singh, D., & Sharma, N. (2022). Comprehensive Study of Microstrip Patch Antenna Using Different Feeding Techniques. ECS Transactions, 107(1), 9545. http://dx.doi.org/10.1149/10701.9545ecst

Khan, A., & Nema, R. (2012). Analysis of five different dielectric substrates on microstrip patch antenna. *International journal of computer applications*,vol.55(14). https://doi.org/10.5120/8826-2905

Kiani, N., Hamedani, F. T., & Rezaei, P. (2021). Polarization controlling plan in graphene-based reconfigurable microstrip patch antenna. *Optik*, 244, 167595. https://doi.org/10.1016/j.ijleo.2021.167595

Kouhalvandi, L., Ceylan, O., Paker, S., & YAĞCI, H. B. (2017). Design and realization of a novel planar array antenna and low power LNA for Ku-band small satellite communications. *Turkish Journal of Electrical Engineering and Computer Sciences*, 25(2), 1394-1403. https://doi.org/10.3906/elk-1509-148

Lee, C. S., Bai, B., Song, Q. R., Wang, Z. Q., & Li, G. F. (2019). Open complementary split-ring resonator sensor for dropping-based liquid dielectric characterization. IEEE Sensors Journal,19(24),11880-11890. https://doi.org/10.1109/JSEN.2019.2938184

Mishra, B., Verma, R. K., Yashwanth, N., & Singh, R. K. (2022). A review on microstrip patch antenna parameters of different geometry and bandwidth enhancement techniques. *International Journal of Microwave and Wireless Technologies*,14(5),652-673. https://doi.org/10.1017/S1759078721001148

Mishra, R. (2016). An overview of microstrip antenna. *HCTL Open International Journal of Technology Innovations and Research (IJTIR)*, vol. 21(2),pp.39-55. http://dx.doi.org/10.5281/zenodo.161524

Rana, M. S., & Rahman, M. M. (2022). Study of microstrip patch antenna for wireless
392

WWW.EJTAS.COM    EJTAS    2024 | VOLUME 2 | NUMBER 1